\input harvmac
\noblackbox
\def\npb#1#2#3{{\it Nucl.\ Phys.} {\bf B#1} (19#2) #3}

\def\plb#1#2#3{{\it Phys.\ Lett.} {\bf B#1} (19#2) #3}
\def\prl#1#2#3{{\it Phys.\ Rev.\ Lett.} {\bf #1} (19#2) #3}

\def\mpla#1#2#3{{\it Mod.\ Phys.\ Lett.} {\bf A#1} (19#2) #3}

\def\cmp#1#2#3{{\it Comm.\ Math.\ Phys.} {\bf #1} (19#2) #3}

\def\endli{\hfill\break}
\def\frac#1#2{{#1 \over #2}}

\def\p{\partial}
\def\semi{\mathrel{\hbox{$\subset\kern-1em\times$}}}
\def\bar#1{\overline{#1}}

\def\hvezdickazectyr{\hbox{$^{+\mkern-13.8mu\mkern-.015mu\times}$}}
\def\CA{{\cal A}}                   
                   \def\CG{{\cal G}}
                   
\def\CL{{\cal L}}                   
\def\CN{{\cal N}}                   
                   
                   \def\CW{{\cal W}}

\def\R{{\bf R}}                     \def\S{{\bf S}}
\def\T{{\bf T}}                     
\def\Z{{\bf Z}}
\Title{\vbox{\baselineskip12pt
\hbox{hep-th/9705055}
\hbox{PUPT-1700}
\hbox{\ }}}
{\vbox{\centerline{MATRIX THEORY AND}
\bigskip
\centerline{HETEROTIC STRINGS ON TORI}}}
\medskip
\centerline{Petr Ho\v rava\footnote{\hvezdickazectyr}{e-mail address: 
horava@puhep1.princeton.edu}}
\medskip
\centerline{\it Joseph Henry Laboratories, Princeton University}
\centerline{\it Jadwin Hall, Princeton, NJ 08544, USA}
\medskip\medskip\medskip
We consider compactifications of the matrix model of M-theory on 
$\S^1/\Z_2\times\T^d$ for $d>0$, and interpret them as orbifolds of the 
supersymmetric $U(N)$ Yang-Mills theory on $\R\times\T^{d+1}$.  The orbifold 
group acts both on the gauge group and on the $\T^{d+1}$, reduces the gauge 
group to $O(N)$ over $1+1$ dimensional fixed-point submanifolds, and breaks 
half of the supersymmetry.  We clarify some puzzling aspects of the gauge 
anomaly cancellation in the presence of space-time Wilson lines; in general, 
the Yang-Mills theory requires certain Chern-Simons couplings to supergravity 
background fields.  We discuss the possibility that D8-branes are present as 
certain matrix configurations in the Yang-Mills theory, and the fundamental 
fermions emerge as zero modes.  Finally, we point out that the correspondence 
between matrix theory and string theory suggests the existence of a multitude 
of non-trivial RG fixed points and dualities in orbifold Yang-Mills theories 
with eight supercharges in various dimensions.  

\Date{April, 1997}
\nref\bfss{T. Banks, W. Fischler, S.H. Shenker and L. Susskind, ``M Theory as 
a Matrix Model: A Conjecture,'' hep-th/9610043.}
\nref\dvv{R. Dijkgraaf, E. Verlinde and H. Verlinde, ``Matrix String Theory,'' 
hep-th/9703030; ``5D Black Holes and Matrix Strings,'' hep-th/9704018.}
\nref\strfrom{T. Banks and N. Seiberg, ``Strings from Matrices,'' 
hep-th/9702187.}
\nref\hw{P. Ho\v rava and E. Witten, ``Heterotic and Type I String Dynamics 
from Eleven Dimensions,'' hep-th/9510209.}
\nref\hweff{P. Ho\v rava and E. Witten, ``Eleven-Dimensional Supergravity on 
a Manifold with Boundary,'' hep-th/9603142.}
\nref\matror{U.H. Danielsson and G. Ferreti, ``The Heterotic Life of the 
D-Particle,'' hep-th/9610082\endli
S. Kachru and E. Silverstein, ``On Gauge Bosons in the Matrix Model Approach 
to M Theory,'' hep-th/9612162\endli
D. Lowe, ``Bound States of Type I$'$ D-Particles and Enhanced Gauge 
Symmetry,'' hep-th/9702006.}
\nref\matrorkr{N.-W. Kim and S.-J. Rey, ``M(atrix) Theory on an Orbifold and 
Twisted Membrane,'' hep-th/9701139.}
\nref\holo{L. Susskind, ``The World as a Hologram,'' hep-th/9409089.}
\nref\ewdua{E. Witten, ``String Theory Dynamics in Various Dimensions,'' 
hep-th/9503124.}
\nref\fvafa{C. Vafa, ``Evidence for F-Theory,'' hep-th/9602022.}
\nref\dosiss{M.R. Douglas, H. Ooguri and S.H. Shenker, ``Issues in M(atrix) 
Theory Compactification,'' hep-th/9702203.}
\nref\fira{W. Fischler and A. Rajaraman, ``M(atrix) String Theory on K3,'' 
hep-th/9704123.}
\nref\fhrs{W. Fischler, E. Halyo, A. Rajaraman and L. Susskind, ``The 
Incredible Shrinking Torus,'' hep-th/9703102.}
\nref\cswo{P. Ho\v rava, ``Chern-Simons Gauge Theory on Orbifolds: Open 
Strings from Three Dimensions,'' {\it J. Geom.\ Phys.} {\bf 21} (1996) 1; 
hep-th/9404101\endli
P. Ho\v rava, ``Orbifold Approach to Open String Theory,'' PhD. Thesis 
(Institute of Physics, Czech Academy of Sciences, Prague; 1991).}
\nref\bamo{T. Banks and L. Motl, ``Heterotic Strings from Matrices,'' 
hep-th/9703218.}
\nref\taylor{W. Taylor IV, ``D-Brane Field Theory on Compact Spaces,'' 
hep-th/96111042.}
\nref\lsdual{L. Susskind, ``T Duality in M(atrix) Theory and S Duality in 
Field Theory,'' hep-th/9611164.}
\nref\grtdual{O.J. Ganor, S. Ramgoolam and W. Taylor IV, ``Branes, Fluxes and 
Duality in M(atrix) Theory,'' hep-th/9611202.}
\nref\matrorm{L. Motl, Quaternions and M(atrix) Theory in Spaces with 
Boundaries,'' hep-th/961298.}
\nref\bss{T. Banks, N. Seiberg and E. Silverstein, ``Zero and One-Dimensional 
Probes with N=8 Supersymmetry,'' hep-th/9703052.}
\nref\lrnew{D. Lowe, ``Heterotic Matrix String Theory,'' hep-th/9704041\endli
S.-J. Rey, ``Heterotic M(atrix) Strings and Their Interactions,'' 
hep-th/9704158.}
\nref\opent{P. Ho\v rava, ``Backround Duality of Open String Models,'' 
\plb{231}{89}{251}\endli
J. Dai, R.G. Leigh and J. Polchinski, ``New Connections Between String 
Theories,'' \mpla{4}{89}{2073}.}
\nref\kmol{D. Kutasov, E. Martinec and M. O'Loughlin, ``Vacua of M-Theory 
and N=2 Strings,'' hep-th/9603116.}
\nref\ewfive{E. Witten, ``Five-Branes and $M$-Theory on an Orbifold,'' 
hep-th/9512219.}
\nref\powi{J. Polchinski and E. Witten, ``Evidence for Heterotic -- Type I 
Duality,'' hep-th/9510169.}
\nref\bwb{M.R. Douglas, ``Branes within Branes,'' hep-th/9512077\endli
M. Green, J.A. Harvey and G. Moore, ``I-Brane Inflow and Anomalous Couplings 
on D-Branes,'' hep-th/9605033\endli
M.B. Green, C.M. Hull and P.K. Townsend, ``D-Brane Wess-Zumino Actions, 
T-Duality and the Cosmological Constant,'' hep-th/9604119\endli
see also: J. Polchinski, ``TASI Lectures on D-Branes,'' hep-th/9611050, 
Section~3.5.}
\nref\fivecs{N. Seiberg,``Five-Dimensional SUSY Field Theories, Non-Trivial 
Fixed Points and String Dynamics,'' hep-th/9608111\endli
D.R. Morrison and N. Seiberg, ``Extremal Transitions and Five-Dimensional 
Supersymmetric Field Theories,'' hep-th/9609070\endli
M.R. Douglas, S. Katz and C. Vafa, ``Small Instantons, Del Pezzo Surfaces and 
Type I$'$ Theory,'' hep-th/9609071\endli
K. Intriligator, D.R. Morrison and N. Seiberg, ``Five-Dimensional 
Supersymmetric Gauge Theories and Degenerations of Calabi-Yau Spaces,'' 
hep-th/9702198.}
\nref\ewflux{E. Witten, ``On Flux Quantization in $M$-Theory and the Effective 
Action,'' hep-th/9609122.}
\nref\dkps{M.R. Douglas, D. Kabat, P. Pouliot and S.H. Shenker, ``D-Branes and 
Short Distances in String Theory,'' hep-th/96088024.}
\nref\mdrev{M.R. Douglas, ``Superstring Dualities, Dirichlet Branes and the 
Small Scale Structure of Space,'' hep-th/9610041.}
\nref\brafrom{T. Banks, N. Seiberg and S. Shenker, ``Branes from Matrices,'' 
hep-th/9612157.}
\nref\eightinst{B. Grossmann, T.W. Kephart and J.D. Stasheff, 
``Solutions to Yang-Mills Field Equations in Eight Dimensions and the Last 
Hopf Map,'' \cmp{96}{84}{431}; {\bf 100} (1985) 311\endli
D.B. Fairlie and J. Nuyts, ``Spherically Symmetric Solutions of Gauge 
Theories in Eight Dimensions,'' {\it J. Phys.} {\bf A17} (1984) 2867\endli
E. Corrigan C. Devchand, D.B. Fairlie and J. Nuyts, ``First-Order Equations 
for Gauge Fields in Spaces of Dimension Greater than Four,'' 
\npb{214}{83}{452}\endli
R.S. Ward, ``Completely Solvable Gauge-Field Equations in Dimension Greater 
than Four,'' \npb{236}{84}{381}.\endli
R. D\"undarer, F. G\"ursey and C.-H. Tze, ``Generalized Vector Products, 
Duality, and Octonionic Identities in $D=8$ Geometry'' 
{\it J. Math.\ Phys.} {\bf 25} 
(1984) 1496; ``Self-Duality and Octonionic Analyticity of $S^7$-Valued 
Antisymmetric Fields in Eight Dimensions,'' \npb{266}{86}{440}\endli
S. Fubini and H. Nicolai, ``The Octonionic Instanton,'' 
\plb{155}{85}{369}\endli
J.A. Harvey and A. Strominger, ``Octonionic Superstring Solitons,'' 
\prl{66}{91}{549}\endli
L. Baulieu, H. Kanno and I.M. Singer, ``Special Quantum Field Theories in 
Eight and Other Dimensions,'' hep-th/970467.}
\nref\fsen{A. Sen, ``$F$-Theory and Orientifolds,'' hep-th/9605150.}
\nref\sethis{S. Sethi and L. Susskind, ``Rotational Invariance in the M(atrix) 
Formulation of Type IIB Theory,'' hep-th/9702101.}
\nref\rozali{M. Rozali, ``Matrix Theory and U-Duality in Seven Dimensions,'' 
hep-th/9702136.}
\nref\berro{M. Berkooz and M. Rozali, ``On Transverse Five-Branes in M(atrix) 
Theory on $T^5$,'' hep-th9704089.}
\newsec{Introduction}

Some of the mystery of M-theory has been recently removed -- 
a very interesting proposal for its non-perturbative description is now 
available \bfss , suggesting that M-theory could be described in terms of a 
matrix model.  This ``matrix theory'' proposal works with microscopic degrees 
of freedom which exhibit the virtues of D-branes that have been so prominent 
in our understanding of non-perturbative string dynamics, and elevates them to 
eleven dimensions.  

The proposal has already passed a whole set of tests, in particular in the 
cases that correspond to compactifications with the maximal amount of 
supersymmetry.  Not only is it apparently possible to reconstruct various 
expected string dualities, but also string perturbation theory in the 
light cone gauge seems to be embedded in the non-perturbative framework of 
matrix theory \refs{\dvv,\strfrom}.  Since the light-cone-gauge perturbation 
expansion of string theory is known to be Lorentz invariant, this gives 
additional credibility to the original matrix theory proposal.  

One of the next steps is to consider compactifications of M-theory that 
respect half of the original supersymmetry, and test whether and how matrix 
theory reproduces the properties of non-perturbative string theory, 
M-theory and F-theory, expected on the basis of various conjectured dualities 
among their vacua.   In this paper, we will take certain steps in this 
direction, and will study compactifications of M-theory on $\S^1/\Z_2\times
\T^d$ in the matrix theory framework.  

These orbifold vacua of M-theory are known to correspond to heterotic string 
theory, which gives some additional motivation for their study in matrix 
theory.  In particular, here are some of the relevant issues: 

\item{(1)}  The relation between M-theory on manifolds with boundaries and 
the $E_8\times E_8$ heterotic string theory \refs{\hw,\hweff} predicts the 
emergence of $E_8$ boundary degrees of freedom in M-theory.  One would want 
to understand better the microscopic origin of these boundary degrees of 
freedom.  In matrix theory, this question is connected with a more general 
question about the BFSS proposal, namely, what is the general prescription 
for compactifications.  One option, suggested already in the original BFSS 
paper \bfss , is that all of the necessary degrees of freedom are already 
present in the matrix theory Hamiltonian in uncompactified eleven dimensions.  
On the other hand, it has been argued in the literature that the proper 
description of the heterotic compactifications requires extra fermions in 
the fundamental representation of the gauge group to be added by hand 
\refs{\matror,\matrorkr}.  

\item{(2)}  The BFSS matrix theory has been argued \bfss\ to exhibit the 
holographic property \holo .  A more detailed picture of holography would 
certainly be useful, as it might elucidate the nature of truly microscopic 
degrees of freedom in matrix theory.  This question also seems to require a 
better understanding of space-time boundaries in M-theory.  

\item{(3)}  Toroidal compactifications of heterotic string theory are 
conjecturally related to various compactifications of string theory, M-theory 
and F-theory \refs{\ewdua,\fvafa} on K3 surfaces.  Such K3 compactifications 
seem to be rather difficult to understand in matrix theory directly 
\refs{\dosiss,\fira}, 
and one might hope that understanding the heterotic side of the conjectured 
dualities in the matrix theory might illuminate some of the difficulties.  

\item{(4)}  It was pointed out in \fhrs\ that matrix theory, in conjuction 
with the expected string dualities in various dimensions, leads to predictions 
about the non-perturbative dynamics of supersymmetric Yang-Mills gauge 
theories in the corresponding dimensions.  In the case of toroidal 
compactifications analyzed in \fhrs , matrix theory typically leads to 
postdictions rather than predictions, as the non-trivial fixed points 
predicted by matrix theory are already known from other considerations. 
We will see in this paper that when applied to the matrix theory 
compactifications with half of the maximal supersymmetry, the logic of \fhrs\ 
leads to a multitude of non-trivial predictions about the RG behavior in 
corresponding Yang-Mills gauge theories.  

In section 2 of this paper, we analyze matrix theory compactified on 
$\S^1/\Z_2\times\T^d$, and demonstrate that it is described by Yang-Mills 
gauge theories on orbifolds, or more precisely, orbifold Yang-Mills gauge 
theories -- the action of the orbifold group on the ``space-time'' of the 
Yang-Mills theory is combined with an action on the gauge group itself.  
A similar class of gauge theories on orbifolds has been studied previously 
in a remotely related context in \cswo .  

In section 3 we clarify certain aspects of the anomaly cancellation mechanism, 
relevant to the description of general compactifications with Wilson lines.  
It is argued that for generic compactifications, the Yang-Mills theory 
contains certain Chern-Simons terms that couple the Yang-Mills degrees of 
freedom to a non-trivial supergravity background.  The presence of this 
background is responsible for the local cancellation of gauge anomalies, and 
is needed for the description of Wilson lines in matrix theory.  

In section 4, we further study the physics of D8-branes (or more precisely, 
longitudinally wrapped 9-branes of M-theory) in matrix theory, and suggest 
that they appear as certain topologically non-trivial Yang-Mills 
configurations.  The extra fermions corresponding to 0-8 strings do not have 
to be added by hand on the basis of anomaly cancellation arguments; rather, 
they appear as zero modes in the D8-brane background.  In this scenario, we 
expect the supergravity background and the Chern-Simons couplings to be 
generated when non-zero modes in the configuration describing the D8-branes 
are integrated out.  

In section 5, we briefly compare the Yang-Mills theory description of matrix 
theory on $\S^1/\Z_2\times \T^d$ with the expected behavior in string theory, 
and point out that this comparison suggests the existence of a multitude of 
non-trivial fixed points in the orbifold supersymmetric Yang-Mills theory on 
$\S^1\times\T^d/\Z_2$.  

While this paper was being written, some closely related results were reported 
in \bamo ; in particular, it was independently noted in \bamo\ that matrix 
theory on $\S^1/\Z_2\times\T^d$ is described by super Yang-Mills theory on 
$\S^1\times\T^d/\Z_2$.  

\newsec{Matrix Theory on $\S^1/\Z_2\times\T^d$ as an Orbifold Yang-Mills 
Theory}

According to the original BFSS proposal \bfss , M-theory in the infinite 
momentum frame is described as a system of $N\rightarrow\infty$ ``partons,'' 
represented by D0-branes as the natural carriers of longitudinal 
momentum.  The dynamics is summarized in the quantum mechanical Lagrangian 
with $U(N)$ gauge symmetry, 
\eqn\eebfsslagr{\CL=\frac{1}{2g^2}\int dt\,\tr\left\{(D_0X)^2+\bar\theta 
D_0\theta-\frac{1}{2}([X^i,X^j])^2-\bar\theta\Gamma_i[X^i,\theta]\right\}.}
Here the transverse space-time coordinates $X^i$, $i=1,\ldots 9$, as well as 
their superpartners $\theta$ (in the ${\bf 16}$ of the transverse 
Lorentz group $SO(9)$) are in the adjoint representation of the gauge group 
$U(N)$.  Their $U(N)$ matrix elements represent the lowest open-string modes 
connecting the $N$ D0-branes, whose role has thus been promoted to eleven 
dimensions.  Our normalizations are such that the coupling constant 
$g=\tilde R^{3/2}/(2\pi\ell_{11}^3)$; here $\tilde R$ is the radius of the 
longitudinal dimension, and $\ell_{11}$ is the eleven-dimensional Planck 
length.  

The kinematical light-cone supersymmetries are:
\eqn\eeorigsym{\eqalign{\delta A_0&=\frac{1}{2}\bar\eta\theta,\cr
\delta_\eta X^i&=\frac{1}{2}\bar\eta\Gamma^i\theta,\cr
\delta_\eta\theta&=-\frac{1}{2}D_0X^i\Gamma_i\eta-\frac{1}{4}[X^i,X^j]
\Gamma_{ij}\eta,\cr}}
and the supersymmetry algebra is closed up to a gauge transformation.  

The $\Z_2$ orbifold vacua of M-theory that we want to understand can be 
conveniently studied as $\Z_2$ orbifolds of matrix theory compactified on a 
torus.  Matrix theory on $\T^d$ is described 
\refs{\bfss,\taylor,\lsdual,\grtdual} by the maximally supersymmetric $U(N)$ 
Yang-Mills gauge theory on the dual torus, $\hat\T^d$.%
\foot{{}From now on, we will drop the hat denoting the dual torus.  In all 
compactifications considered in this paper, the space-like part of the 
Yang-Mills parameter space is always related to the compact dimensions in 
space-time by T-duality.}
This result can be easily derived from the original BFSS model 
\refs{\bfss,\taylor}.   $\T^d$ is 
a $\Z^d$ orbifold of $\R^d$, and a configuration of $k$ D0-branes on $\T^d$ 
can be represented as an infinite array of D0-branes on $\R^d$ that respect  
the $\Z^d$ symmetry.   The orbifold group $\Z^d$ is represented on the 
original $U(N)$ degrees of freedom of \eebfsslagr\ in terms of commuting 
unitary matrices $U_m$, $m=1,\ldots d$, and the fields of \eebfsslagr\ are 
required to satisfy the corresponding periodicity conditions.  The 
periodicity of the torus along the $m$-th direction is implemented in the 
following condition, 
\eqn\eetransrep{\quad X^i= U_mX^iU_m^{-1}+e^i_m,}
with $e^i_m$ the components of the corresponding lattice vector.  
This condition is solved by 
\eqn\eexder{X_i= D_i\equiv\p_i+A_i,}
where $D_i$ is the covariant derivative along an extra dimension of the 
Yang-Mills theory.  \eexder\ is of course nothing but the traditional formula 
expressing the action of T-duality on the D-brane world-volume fields.  This 
interpretation invokes the extrapolation of certain aspects of the physics of 
slowly moving D0-branes in weakly coupled string theory, a useful heuristic 
analogy that will play its role below.  

Now we want to describe $\Z_2$ orbifolds of matrix theory compactifications 
on $\T^{d+1}$, where the orbifold group reflects one of the dimensions in 
$\T^{d+1}$.%
\foot{In this paper, we always choose the $\Z_2$ orbifold action such that 
the longitudinal direction of matrix theory is parallel to the space-time 
boundary.}
One could start from the original Lagrangian \eebfsslagr , 
and implement the orbifold procedure which contains both the $\Z^d$ that 
defines the torus and the $\Z_2$ which acts on one of the dimensions of the 
torus by reflection.  Due to this non-trivial action of $\Z_2$ on the torus, 
the full orbifold group is a product of the non-abelian semi-direct product 
$\Z_2\semi\Z$ with the abelian group $\Z^d$ representing the ``transverse'' 
torus, 
\eqn\eeorbgr{\CG=(\Z_2\semi\Z)\times\Z^d.}
One can now follow the steps performed in the case of the toroidal 
compactification, and factorize the BFSS theory by the non-Abelian discrete 
group \eeorbgr .  

Alternatively, one can perform the orbifold procedure in two steps, and 
consider $\Z_2$ orbifolds of the $d+2$ dimensional $U(N)$ Yang-Mills theory 
that describes matrix theory compactified on $\T^{d+1}$.  In this paper, we 
will mostly follow the latter strategy.  First, however, let us make the 
``one-step'' approach a little more explicit (for a related discussion, see 
\refs{\matrorm,\matrorkr,\matror}).  

The translations by elements of $\Z^{d+1}$ have been represented by the 
commuting unitary matrices $U_m$.  To fully specify the orbifold group 
\eeorbgr\ action in matrix theory, we choose a unitary matrix $\Omega$ that 
represents the generator of $\Z_2$.  One can use the extrapolation from 
weakly coupling string theory to argue that $\Omega$ will act on the $U(N)$ 
adjoint fields by a matrix transposition.  In Type IIA theory, the $\Z_2$ of 
interest is an orientifold symmetry; in particular, it changes the 
orientation of all open strings.  In matrix theory, the $U(N)$ matrix elements 
correspond to the lowest states of open strings stretching between D0-branes, 
and the change of world-sheet orientation will transpose the $U(N)$ matrices. 
 Thus, we define 
\eqn\eeactommat{\Omega M\Omega =M^{T}}
for any matrix $M$ in the adjoint of $U(N)$.  The full action of $\Omega$ on 
the fields of the Yang-Mills multiplet in addition contains the space-time 
$\Z_2$ reflection 
\eqn\eerefff{X^1\rightarrow -X^1,}
and combines with the representation $U_{m}$ of the torus translations 
\eetransrep\ into a representation of the orbifold group \eeorbgr .  

Now, consider the spinor fields.  On the fermions of eleven-dimensional 
M-theory, the $\Z_2$ reflection \eerefff\ of one of the transverse 
coordinates acts by \hw 
\eqn\eeorelef{\xi\rightarrow \Gamma_1\xi.}  
In the light cone gauge, we impose 
\eqn\eelcgelef{\Gamma^+\xi=0,}
which reduces the 32-component spinor $\xi$ to a 16-component one that we 
will denote $\eta$.  {}We can further choose our gamma matrices such that 
$\Gamma_1=\Gamma_2\ldots\Gamma_{10}\Gamma_0=\equiv\frac{1}{2}\Gamma_2\ldots
\Gamma_9[\Gamma_+,\Gamma_-]$, so that the orbifold action \eeorelef\ in the 
light cone frame becomes
\eqn\eefinorcfer{\eta\rightarrow\Gamma_2\ldots\Gamma_9\eta,}
i.e.\ the fermions are chiral in the transverse eight-dimensional space at 
the boundary.  On the fields of the Yang-Mills theory, which are obtained by 
dimensionally reducing the ten-dimensional $\CN=1$ theory, this can be 
rewritten as 
\eqn\eecccymf{\eta\rightarrow\Gamma_{01}\eta.}
In terms of the full Yang-Mills field content, we end up with the following 
orbifold conditions, 
\eqn\eeactorbf{\eqalign{X_1&=-\Omega\,X_1\,\Omega,\cr
A_0&=-\Omega\,A_0\,\Omega,\cr}
\qquad\eqalign{X_i&=\Omega\,X_i\,\Omega,\quad i=2,\ldots,9,\cr
\theta&=-\Omega\,(\Gamma_{01}\theta)\,\Omega.\cr
}}
The action of the orbifold group on the fields induces an action on the 
symmetries of the original Lagrangian: 
\eqn\eeactorbsymm{\eqalign{Q&\rightarrow\Gamma_{01}Q,\cr
G&\rightarrow-\Omega\,G\,\Omega.\cr}}
Only a subgroup of the original symmetry group will survive the 
projection; in particular, only one half of the original supersymmetry is 
preserved in the orbifold theory, and the $U(N)$ gauge symmetry is reduced to 
$O(N)$.  

So far we have projected the theory to its $\Z_2$-invariant subsector.  Since 
the orbifold group does not act freely, we might expect that self-consistency 
will require some extra degrees of freedom corresponding to ``twisted 
sectors.'' In \refs{\matror,\matrorkr} it was indeed argued that sixteen 
fermions in the fundamental representation of the gauge group have to be 
added; the precise structure of these twisted modes has been inferred from 
the extrapolation of D0-brane dynamics in weakly coupled Type IA theory, 
where they correspond to the massless modes of the 0-8 strings in the Ramond 
sector.  

\subsec{Matrix Theory on $\S^1/\Z_2$}

Even though the main focus of this paper is on compactifications on 
$\S^1/\Z_2\times\T^d$ for $d>0$, we first review the simple case of $d=0$ 
to set the stage for more general compactifications.  
More details on the $d=0$ case can be found in \refs{\matror,\matrorkr,\bss}.%
\foot{The case of matrix theory on $\S^1/\Z_2$ has been further analyzed 
in considerable detail in recent papers, \refs{\bamo,\lrnew}.}

This compactification of matrix theory is described by $1+1$ dimensional 
Yang-Mills theory whose gauge group has been reduced by the $\Z_2$ orbifold 
condition \eeactorbf\ to $O(N)$; half of the sixteen original supersymmetries 
are broken, and the theory enjoys $(0,8)$ supersymmetry in $1+1$ dimensions.  
The field content of the original $U(N)$ theory has been reduced to 
\eqn\eefields{A_{0,1}^{[IJ]},\ \ X_i^{(IJ)},\ \ x_i,\ \ \theta_\alpha^{[IJ]},
\ \ \theta_{\dot\alpha}^{(IJ)},\ \ \vartheta_{\dot\alpha}.}
Here $(IJ)$ indicates the symmetric traceless representation of $O(N)$, and 
$x_i,\vartheta_{\dot\alpha}$ are the singlet parts of the corresponding $U(N)$ 
fields; of course, the $1+1$ chirality of the fermions is correlated with 
their space-time chirality.  

The Lagrangian of this theory is again that of the dimensionally reduced 
Yang-Mills theory, with the gauge group representations appropriately reduced 
from $U(N)$ to $O(N)$, 
\eqn\eelagtwo{\CL=\frac{1}{g^2}\int_{\S^1\times\R}d^2\sigma\,\tr\left(
-\frac{1}{4}F^2+\frac{1}{2}(D_\mu X^i)^2-\frac{1}{4}([X^i,X^j])^2+
\frac{1}{2}(\p_\mu x^i)^2+{\rm fermions}\right).}
The gauge coupling constant $g^2$ and the radius $\rho_1$ of the $\S^1$ in the 
Yang-Mills theory are related to the radius $R_1$ of the space-time orbifold 
dimension $\S^1/\Z_2$ by 
\eqn\eecouprelst{g^2=\frac{\tilde R^2}{R_1\ell_{11}^3},\qquad \rho_1=(2\pi)^2
\frac{\ell_{11}^3}{\tilde RR_1}.}

The surviving field content \eefields\ is chiral, and \eelagtwo\ represents an 
anomalous theory.  Under an $O(N)$ gauge transformation, the effective action 
transforms as 
\eqn\eeanomfirs{\delta\CW=32\cdot\frac{1}{2\pi}\int\tr\;(\epsilon F).}
It has been argued \refs{\bss,\matrorkr} on precisely this basis that a set 
of 32 chiral fermions $\chi_r^I$ in the fundamental representation of 
the gauge group $O(N)$ should be added to cancel the anomaly.  Their 
contribution to \eelagtwo\ is given by
\eqn\eefermilag{\sum_{r=1}^{32}\int d^2\sigma\;\left(\chi^I_r\p_+\chi^I_r+
\chi^I_rA_+^{IJ}\chi^J_r\right),}
and their contribution to $\delta\CW$ precisely cancels that of \eeanomfirs .  
It is the origin of these extra fields that we would want to understand 
better from first principles in matrix theory; we will return to this question 
in section~4.  

Thus, compactification of matrix theory on $\S^1/\Z_2$ leads to a relatively 
standard, 
Lorentz invariant $(0,8)$ supersymmetric field theory in $1+1$ dimensions.  
Most of the field content was determined by the $\Z_2$ orbifold projection, 
and the rest had to be added in order to cancel the $1+1$ dimensional gauge 
anomaly of those fields that survived the $\Z_2$ projection.  

Now we will see that the fun actually begins in $2+1$ dimensions.  

\subsec{Matrix Theory on $\S^1/\Z_2\times\S^1$}

This vacuum is a $\Z_2$ orbifold of the compactification of matrix theory 
on $\T^2$, and should therefore be describable as an orbifold of the $2+1$ 
dimensional $U(N)$ gauge theory on the dual $\T^2$.  

This orbifold theory cannot be the naive, $2+1$ Lorentz-invariant $O(N)$ gauge 
theory on $\T^2$ with $\CN=4$ supersymmetry.  One can give several arguments 
that support this statement:  

\item{(1)} In the limit where the space-time $\S^1$ decompactifies, the 
Yang-Mills theory undergoes dimensional reduction to $1+1$ dimensions, and 
should reproduce the $1+1$ dimensional theory reviewed in the previous 
subsection.  In particular, one of the scalars of the $1+1$ dimensional 
Yang-Mills multiplet should come from the third component of the vector 
potential in $2+1$ dimensions.  However, all scalars that survive the $\Z_2$ 
projection in $1+1$ dimensions are in the wrong representation (either 
symmetric traceless or scalar) of $O(N)$ to become the third component of an 
$O(N)$ gauge field in $2+1$ dimensions.  

\item{(2)}  In the same limit, the reduced Yang-Mills theory exhibits $(0,8)$ 
supersymmetry in $1+1$ dimensions.  The usual Poincar\'e invariant gauge 
theory in $2+1$ dimensions would give, upon dimensional reduction, a theory 
with $(4,4)$ supersymmetry. 

\item{(3)} A closely related space-time argument also indicates that the 
theory cannot exhibit full $2+1$ dimensional Lorentz invariance.  In the 
limit where the space-time $\S^1/\Z_2\times\S^1$ 
shrinks to zero volume, we expect matrix theory on $\S^1/\Z_2\times\S^1$ to 
describe the Type I and heterotic $SO(32)$ string theory.  These string vacua 
are super Poincar\'e invariant in ten dimensions, and the matrix theory 
description should exhibit at least an $SO(7)$ subgroup of the transverse 
$SO(8)$ Lorentz symmetry manifest.  In the corresponding compactifications on 
$\T^2$, the $SO(9,1)$ symmetry of the Yang-Mills theory decomposes as 
$SO(7)\times SO(2,1)$, where $SO(7)$ is the Lorentz group in the transverse 
dimensions, and $SO(2,1)$ is the Yang-Mills Lorentz group.  The sixteen 
supercharges, in the ${\bf 16}$ of $SO(9,1)$, decompose as 
$({\bf 8},{\bf 2})$.  Since we expect the $\Z_2$ orbifold to preserve the 
$SO(7)$ symmetry, it must act trivially on the ${\bf 8}$.  However, the 
orbifold should break half of the supersymmetry, so it has to project out 
half of the ${\bf 2}$ of the $2+1$ dimensional Lorentz group.  This 
is of course incompatible with full $2+1$ dimensional Lorentz invariance.  

The correct answer, in agreement with all these observations, can be inferred 
from ``first principles'' by implementing the orbifold construction directly 
on the original $0+1$ dimensional theory \eebfsslagr .  In terms of the $2+1$ 
dimensional Yang-Mills gauge theory, the orbifold group $\Z_2$ will act not 
only as a matrix transposition on the $U(N)$ adjoints, and simultaneously as 
a reflection of the third dimension in the Yang-Mills parameter space,
\eqn\eesimga{\Omega(\sigma^0,\sigma^1,\sigma^2)=(\sigma^0,\sigma^1,-\sigma^2).}
On the Yang-Mills one-form $A\equiv A_\mu d\sigma^\mu$ and the rest of 
the multiplet, the $\Z_2$ will act by 
\eqn\eeorbabst{\Omega:\qquad\left\{
\eqalign{A(\sigma^\mu)&\ \rightarrow\ -\Omega\,A(\Omega(
\sigma^\mu))\,\Omega,\cr
X^i(\sigma^\mu)&\ \rightarrow\ \Omega\,X^i(\Omega(\sigma^\mu))
\,\Omega,\cr
\theta(\sigma^\mu)&\ \rightarrow\ -\Omega\,(\Gamma_{01}\theta
(\Omega(\sigma^\mu)))\,\Omega.\cr}\right.}
In particular, the components of the gauge field that survive the projection 
satisfy 
\eqn\eeorbfield{\eqalign{A_{0,1}^{IJ}(\sigma^0,\sigma^1,\sigma^2)&=
-A_{0,1}^{JI}(\sigma^0,\sigma^1,-\sigma^2),\cr
A_2^{IJ}(\sigma^0,\sigma^1,\sigma^2)&= A_2^{JI}(\sigma^0,\sigma^1,-\sigma^2),
\cr}}
in accord with remark~(1) above.  

These conditions define a certain Yang-Mills theory on $\S^1\times\S^1/\Z_2$.  
More precisely, they define an ``orbifold Yang-Mills gauge theory,'' since 
the orbifold group acts simultaneously on the underlying manifold $\T^2$ as 
well as the internal $U(N)$ degrees of freedom.  
Alternatively, this $2+1$ dimensional gauge theory can be viewed as a theory 
on a manifold with two $1+1$ dimensional boundaries, and the boundary 
conditions determined by the orbifold conditions \eeorbfield .  It is 
described by the $9+1$ dimensional $\CN=1$ Yang-Mills Lagrangian, 
dimensionally reduced to $2+1$; in particular, notice that the theory still 
exhibits all sixteen supersymmetries in the bulk, broken to eight 
supersymmetries only at the boundaries.  

Even though our theory is a gauge theory in $2+1$ dimensions, its gauge 
symmetry is potentially anomalous.  The mechanism is precisely analogous to 
the emergence of gravitational anomaly in eleven-dimensional supergravity on 
a manifold with boundaries \refs{\hw,\hweff}.  Due to the chiral boundary 
conditions induced by the orbifold at the $1+1$ dimensional boundaries, the 
effective action exhibits an anomaly supported at the boundaries.  Under a 
gauge transformation $\epsilon(\sigma^\mu)$ that respects the $\Z_2$ symmetry, 
the effective action transforms as follows, 
\eqn\eeeffanom{\delta\CW=16\cdot\frac{1}{2\pi}\int_{V_1}\tr\;(\epsilon_1 F)+
16\cdot\frac{1}{2\pi}\int_{V_2}\tr\;(\epsilon_2 F).}
Here $V_1,V_2$ are the two $1+1$ dimensional components of the boundary, 
while $\epsilon_1=\epsilon|_{V_1}$ and $\epsilon_2=\epsilon|_{V_2}$ have been 
used to denote the gauge transformation $\epsilon$ evaluated at the two 
boundary components.  

The overall coefficient in \eeeffanom\ has been determined as follows.  
Under gauge transformations that project to $1+1$ dimensions, i.e.\ are 
independent of the coordinate $\sigma^2$ normal to the boundaries, the total 
anomaly should equal the anomaly in the dimensionally reduced $1+1$ 
dimensional theory; in the $2+1$ dimensional theory, therefore, each boundary 
component supports one half of the $1+1$ dimensional gauge anomaly 
\eeanomfirs .  

This anomaly can be canceled by adding chiral degrees of freedom at each 
boundary component.  A natural choice would be to split the fermions of 
\eefermilag\ into two groups of sixteen, and place each group at one component 
of the boundary.  We will see in sections~3 and~4 that the actual 
anomaly-cancellation mechanism is a little more sophisticated.  

\subsec{Matrix Theory on $\S^1/\Z_2\times\T^d$}

Now we can easily generalize our description to $d>1$.  To simplify our 
discussion, we assume throughout the paper that $\S^1/\Z_2\times\T^d$ carries 
a rectangular metric, and we also set all theta angles corresponding to 
antisymmetric background fields to zero; generalizations to generic tori are 
straightforward.  

The arguments of the previous subsection can be repeated to show that each 
additional dimension compactified on a space-time circle gives rise to 
a dimension in the Yang-Mills parameter space which is odd under 
the $\Z_2$ action.  We will denote the corresponding coordinates by 
$\sigma^\mu=(\sigma^0,\sigma^1,\ldots,\sigma^{d+1})$. The orbifold group  
acts on the Yang-Mills parameter space by $\Omega$, 
\eqn\eeactcoords{\Omega:\quad\left\{\eqalign{\sigma^\mu&\rightarrow\sigma^\mu,
\quad\ \ \mu=0,1,\cr\sigma^\mu&\rightarrow -\sigma^\mu,\quad\mu=2,\ldots, d+1.
\cr}\right.}
The singular locus of this $\Z_2$ action has $2^d$ components, each of them 
being a $1+1$ dimensional cylinder $\S^1\times\R$.  The full action of the 
orbifold group on the Yang-Mills multiplets is again given by \eeorbabst .  

Thus, we obtain the following correspondence, 
\eqn\eecorrw{({\rm matrix\ theory\ on}\ \S^1/\Z_2\times\T^d)\ 
\longleftrightarrow\ ({\rm Yang\ Mills\ theory\ on}\ \S^1\times\T^d/\Z_2).}
In this correspondence, the supersymmetric Yang-Mills theory on 
$\S^1\times\T^d/\Z_2$ is such that the orbifold group acts simultaneously on 
the Yang-Mills parameter space and on the gauge group representations.  
The theory is the maximally supersymmetric, $U(N)$ Yang-Mills gauge theory in 
the bulk, and is described by the dimensionally reduced Yang-Mills Lagrangian, 
\eqn\eefinlaf{\CL=\frac{1}{g^2}\int d^{d+2}\sigma\,\tr\left(
-\frac{1}{4}F^2+\frac{1}{2}(D_\mu X^i)^2-\frac{1}{4}([X^i,X^j])^2+
{\rm fermions}\right).}
Over the $1+1$ dimensional manifolds of fixed points, the gauge symmetry is 
reduced to $O(N)$, and one half of the sixteen supersymmetries are broken.  
{}From the point of view of the $1+1$ dimensional singular locus, the boundary 
conditions are chiral; fields that are even under the $\Z_2$ symmetry are 
precisely the multiplets \eefields\ of the $1+1$ dimensional theory, and the 
eight unbroken supersymmetries all carry the same $1+1$ dimensional 
chirality.  

Several comments are in order:

\item{(1)} It is of course not surprising to see the role of $\S^1$ and 
$\S^1/\Z_2$ interchanged as we switch from the space-time to the Yang-Mills 
description of the theory according to \eecorrw .  These two descriptions are 
related by T-duality, and it is a well-known property of T-duality in open 
string models that it exchanges $\S^1$ with $\S^1/\Z_2$ \opent .%
\foot{The orbifold geometry $\S^1\times\T^d/\Z_2$ has also been encountered 
in a similar context by Kutasov, Martinec and O'Loughlin in their approach to 
non-perturbative M-theory \kmol .  I am grateful to Martin O'Loughlin for 
discussions on the results of \kmol .}

\item{(2)} Gauge theories of this orbifold type are not really new; their 
close relatives were studied in a context related to D-branes and to open 
string vacua in \cswo .  It was shown in \cswo\ that Chern-Simons gauge 
theories on $2+1$ dimensional orbifolds are related to conformal field 
theories on surfaces with boundaries and/or crosscaps, and provide a natural 
framework for the study of orientifold and D-brane conformal field theories.  
The orbifold gauge theories of \cswo\ exhibit some striking similarities with 
the theories we encounter in the matrix model correspondence \eecorrw .

\item{(3)} The Yang-Mills coupling constant $g^2$ and the radii $\rho_m$ 
of the Yang-Mills parameter space $\S^1\times\T^d/\Z_2$ are related to the 
radii $R_m$ of the space-time $\S^1/\Z_2\times\T^d$ by 
\eqn\eegaucoup{g^2=(2\pi)^{2d}\;\ell_{11}^{3d-3}\frac{\tilde R^{2-d}}{R_1
\ldots R_{d+1}},\qquad \rho_m=(2\pi)^2\frac{\ell_{11}^3}{\tilde RR_m}.}
In cases other than $d=2$, the coupling constant $g^2$ is dimensionful, and 
can be rescaled to give 
\eqn\eedlesscc{\tilde g^2=\frac{g^2}{(\rho_1\ldots\rho_{d+1})^{(d-2)/(d+1)}}
=(2\pi)^4\frac{\ell_{11}^3}{(R_1\ldots R_{d+1})^{3/(d+1)}}.}

Let us return to the orbifold Yang-Mills theory of \eecorrw .  This theory is 
potentially anomalous, with the anomaly supported at the planes of fixed 
points.  Just as in the case of $d=1$, we find that under a gauge 
trasformation, the effective action transforms as follows,
\eqn\eeefftranhigh{\delta\CW=\frac{1}{2^d}\cdot\frac{16}{\pi}
\sum_\alpha\int_{V_\alpha}\tr\;(\epsilon F).}
(Here $\alpha=1,\ldots,2^d$ parametrizes the components of the singular 
locus.)  
The overall coefficient has again been determined by symmetry arguments 
similar to those of \refs{\hw,\hweff}; each plane of fixed points supports 
an anomaly equal to $2^{-d}$ times the anomaly of the $1+1$ dimensional 
theory.  We can try to cancel this anomaly by adding chiral degrees of 
freedom to the theory, the obvious choice being a set of fermions described by 
\eqn\eefermih{\sum_{\alpha}\int_{V_\alpha}d^2\sigma\;\tr\;
\chi(\p_++A_+)\chi,}
a prescription which can only work if the number of space-time dimensions 
compactified on $\T^d$ is lower than six.  Another troubling situation 
would occur in the presence of space-time Wilson lines on $\T^d$ for any $d$; 
the natural Yang-Mills description of such vacua would require the fermions to 
be located away from $V_\alpha$, and anomalies could not be cancelled 
locally by the suggested mechanism.  These arguments indicate that our 
picture of anomaly cancellation still misses some crucial ingredients.  We 
will present a solution to this puzzle in section~3.  

\subsec{Comparison to Weak String Coupling}

Resorting once again to the extrapolation into the regime of slowly moving 
D-branes in weakly coupled string theory, we can argue that precisely this 
orbifold Yang-Mills theory could have been expected to appear in the 
description of matrix theory on $\S^1/\Z_2\times \T^d$.  

Consider a system of $N$ D0-branes in Type IA string theory on a torus, close 
to one of the orientifold planes.  
In order to keep the string coupling small everywhere, we put sixteen 
D8-branes on top of each orientifold plane.  By T-duality in the orbifold 
dimension, this system is mapped to a system of $N$ D1-branes in Type I 
theory; the orientifold plane becomes space-filling, and intersects each 
D1-string along the $1+1$ dimensional manifold which is the string itself.  
By another T-duality, now in one of the dimensions parallel to the original 
D8-branes, this configuration is again mapped to Type IA theory, where it 
represents $N$ D2-branes ending on two $8+1$ dimensional orientifold planes.  
This can be further dualized along any of the extra 
dimensions parallel to the original D8-branes, leading to a hierarchy of 
systems in successive dimensions representing $N$ D$p$-branes 
intersecting $2^{p-1}$ orientifold planes of dimension $11-p$ along $2^{p-1}$ 
strings.  The Yang-Mills description of matrix theory on $\S^1/\Z_2\times\T^d$ 
in terms of the orbifold gauge theory corresponds to the world-volume gauge 
theory of precisely this configuration of branes intersecting orientifold 
planes (and the D$(10-p)$-branes T-dual to the original 
D8-branes) for $d={p-1}$, and extrapolated into the regime of strong string 
coupling.  

\newsec{Anomaly Cancellation and Chern-Simons Terms in Matrix Theory}

The simple matrix theory description of compactifications on $\S^1/\Z_2\times
\T^d$ that we obtained in the previous section seems to run into problems in 
the cases that do not permit local cancellation of anomalies by distributing 
the fermions among the components of the singular locus.  Such configurations 
would appear for $d$ higher than five, but also in any dimension in the 
presence of generic Wilson lines around the space-time torus $\T^d$. 

In this section, we present a resolution which is very close in spirit to the 
resolution of similar puzzles in orientifold compactifications of string 
theory and M-theory, cf.\ e.g.\ \ewfive .  While this argument will indeed 
show that the theory is non-anomalous, we will see indications that the 
suggested resolution should perhaps be interpreted as an effective description 
of a more microscopic mechanism; we will indeed suggest a more microscopic 
picture in section~4.  

In the previous section, we have seen that the orbifold Yang-Mills theory, 
when extrapolated to the regime of weak string coupling, corresponds to the 
world-volume gauge theory of branes intersecting other branes and orientifold 
planes along strings.  In that extrapolated context, it is known that no 
anomalies appear \refs{\powi,\ewfive,\bwb}; the gauge theory on the brane 
world-volume contains a topological coupling to the RR background fields 
\bwb , 
\eqn\eecstot{\int C\wedge\tr\,\, e^F,}
and the orientifold planes (as well as the branes that intersect the 
world-volume) carry a non-zero RR charge.  Both the orientifold planes and the 
intersecting branes are separately non-anomalous, because the would-be 
anomalies are cancelled locally due to the anomaly inflow from the bulk 
mediated by \eecstot .  

Assuming no surprises happen as we extrapolate to the regime of strong string 
coupling, this mechanism resolves our puzzle, at the cost of introducing the 
explicit Chern-Simons couplings to $\sigma$-dependent RR backgrounds.%
\foot{Even though these background fields now belong to the multiplet of 
eleven-dimensional supergravity, we will frequently refer to them using 
the ten-dimensional string language.}
We postulate that the fixed-point planes $V_\alpha$ carry a RR charge, and 
add the explicit Chern-Simons terms \eecstot\ to the matrix theory Lagrangian; 
the local anomaly cancellation is then ensured by the anomaly inflow from the 
bulk of the Yang-Mills parameter space.  

Let us study the suggested mechanism in more detail.  Since our focus is 
on infrared effects, it is natural to consider the $9+1$ dimensional 
Yang-Mills theory; the cases describing uncompactified space-time dimensions 
can be recovered from this theory by dimensional reduction.  
In this context, each component $V_\alpha$ of the fixed-point locus is an 
electric source for a RR 2-form $C'$, and a magnetic source for 
the RR 6-form $C$.  In particular, $C$ will satisfy the modified Bianchi 
identity, 
\eqn\eembi{dG=e_\alpha\delta_{V_\alpha},}
where $G$ is the field strength of $C$ and $\delta_{V_\alpha}$ is the delta 
function localized at the $1+1$ dimensional submanifold $V_\alpha$ 
(and interpreted as a $d$-form), and $e_\alpha$ is the charge of $V_\alpha$.  
This charge can be determined by extrapolating from the weakly-coupled regime 
\refs{\bwb,\ewfive}; in the case that corresponds to matrix theory on 
$\S^1/\Z_2\times\T^d$, each individual component $V_\alpha$ of the singular 
locus carries $-2^{5-d}$ units of the corresponding D-brane charge.  

The relevant part of \eecstot\ is then 
\eqn\eecstone{-\int G\wedge \tr\;\omega_3(A),}
with the Chern-Simons three-form defined by $d\omega_3(A)=F\wedge F$, and
\eqn\eecsttwo{-\int {}^\star G\wedge\tr\,\omega_7(A),}
where the Chern-Simons seven-form satisfies $d\omega_7(A)=F\wedge F\wedge F
\wedge F$. The first term modifies the anomaly structure of the Yang-Mills 
gauge symmetry, and is responsible for the local anomaly cancellation near 
the fixed-point submanifolds $V_\alpha$.  Under a gauge transformation, we 
find 
\eqn\eeofcour{\eqalign{\delta\left(-\int G\wedge\tr\;\omega_3(A)\right)&=-
\int G\wedge\,\tr\;(D\epsilon\wedge F)\cr
&\qquad{}=\int dG\wedge\tr\;(\epsilon F)=\sum_\alpha e_\alpha
\int_{V_\alpha}\tr\;(\epsilon F),\cr}}
which is precisely the result we need to cancel the anomalies locally at 
$V_\alpha$.  

The second Chern-Simons term, \eecsttwo , measures the D8-brane 
charge carried by the corresponding configuration.  There are two sources 
of the D8-brane charge in the theory: the fixed point submanifolds $V_\alpha$ 
that correspond to the orientifold planes, and the locations of the fermions 
$\chi$ corresponding to the D8-branes themselves.  To ensure the global 
existence of $G$, the total background charge must add up to zero.  Even 
though the fermions are no longer needed for anomaly cancellation, their 
presence is still required by the background charge conservation.  

The configurations with a non-democratic distribution of fermions among the 
fixed-point planes -- corresponding to vacua with generic space-time Wilson 
lines -- are not anomalous; they do require, however, a non-zero RR 
background.%
\foot{Notice that in general, the background breaks the original sixteen 
supercharges to eight supersymmetries even in the bulk.}
Once there is a non-trivial RR background in the theory, we should expect 
a non-trivial NS background to accompany it, e.g.\ on the basis of the 
space-time equations of motion.  It would be very important to 
understand directly in the Yang-Mills theory how these background fields 
self-adjust in order to represent consistent backgrounds of matrix theory with 
the right amount of supersymmetry.  
Given a consistent background, the Chern-Simons terms \eecstot\ will receive 
modifications due to the non-zero curvature and the NS two-form.  The complete 
Chern-Simons couplings in the extrapolated regime of D-brane world-volume 
field theories at weak string coupling can be found in \bwb ; in particular, 
the presence of non-zero curvarure will modify \eecstot\ to \bwb
\eqn\eecsmodif{\int C\wedge\tr\;e^{F}\wedge\sqrt{\hat A(R)}.}

Thus, the requirement of local anomaly cancellation leads us to an 
intricate picture, whereby generic vacua of matrix theory on 
$\S^1/\Z_2\times\T^d$ are described by orbifold Yang-Mills gauge theories with 
couplings to a non-trivial supergravity background.  

\subsec{Wilson Lines}

We have argued that the Yang-Mills theory is locally non-anomalous in the 
vicinity of each plane of fixed points, no matter whether this plane supports 
any of the fermions $\chi$.  Turning this argument around, the fermions do not 
have to be located at the fixed-point planes to cancel anomalies -- we can 
group them in sixteen pairs, and each pair can be supported by a separate 
$1+1$ 
dimensional surface $S_{\tilde\alpha}$ ($\tilde\alpha=1,\ldots 16$) parallel 
to the fixed-point planes $V_\alpha$.  Now we will argue that this 
configuration describes matrix theory on $\S^1/\Z_2\times\T^d$ with generic 
Wilson lines around the $\T^d$; the locations of $S_{\tilde\alpha}$ are given 
by the eigenvalues of the space-time Wilson lines.  

Consider M-theory on $\S^1/\Z_2\times\T^d$ in the presence of space-time 
Wilson lines.  Introducing the following notation,
\eqn\eenotdee{D(\theta)\equiv\pmatrix{\cos\theta&\sin\theta\cr-\sin\theta&
\cos\theta},}
the Wilson lines $W_m$ ($m=1,\ldots d$) can always be brought into the 
following diagonal form, 
\eqn\eewilson{W_m={\rm diag}\,(D(\theta_{i,1}),\ldots D(\theta_{i,16})).}
In the Yang-Mills theory, this configuration is described as follows.  
First we group the 32 fermions into sixteen pairs indexed by 
$\tilde\alpha=1,\ldots 16$.  Each pair will be propagating on a $1+1$ 
dimensional surface $S_{\tilde\alpha}$ which is parallel to the fixed-point 
surfaces $V_\alpha$, and located in the space transverse to $V_\alpha$ at 
coordinates 
\eqn\eelocats{(\sigma^2,\ldots,\sigma^{d+1})=(\pi\theta_{1,\tilde\alpha}\rho_1,
\ldots,\pi\theta_{d,\tilde\alpha}\rho_d).}
As long as the surfaces $S_{\tilde\alpha}$ that support the fermions are 
located away from the fixed-point surfaces $V_\alpha$, the fermions carry the 
fundamental representation of $U(N)$.  Since the fermions are chiral, the 
path integral develops a gauge anomaly, now localized at the $1+1$ 
dimensional surfaces $S_{\tilde\alpha}$.  This anomaly is cancelled by the 
mechanism of section~3.  Recall that each $S_{\tilde\alpha}$ carries 
one unit of the D8-brane charge; therefore, $S_{\tilde\alpha}$ contributes to 
the right hand side of the Bianchi identity \eembi , and the anomaly inflow 
due to the Chern-Simons coupling ensures that the anomaly cancels locally.  
Notice that at $S_{\tilde\alpha}$, half of the original supersymmetry is 
broken.  

The cases with the democratic distribution of fermions among $V_\alpha$ are 
only possible for $d\leq 5$, and correspond to the Wilson lines that break the 
gauge group to $2^d$ copies of $SO(2^{5-d})$.  The generic cases are also 
non-anomalous, but they require a non-zero RR background;  this in turn 
generates a non-trivial gravitational background, and changes the simple 
picture of parallel $1+1$ dimensional surfaces on a flat orbifold.  

Similarly, the vacua with generic locations of D8-branes along the space-time 
orbifold dimension $\S^1/\Z_2$ can be understood in matrix theory as follows.  
The locations of D8-branes in $\S^1/\Z_2$ translate into the Wilson lines in 
the D8-brane gauge group around the $\S^1$ dimension of the Yang-Mills 
parameter space.  This construction requires an explicit coupling to the 
D8-brane gauge fields, and generates non-trivial holonomies for the 
corresponding fermions around the compact dimension of $S_{\tilde\alpha}$.  
The phases of these holonomies are equal to the locations of the corresponding 
D8-branes along the space-time orbifold dimension in the units of its radius.  

\subsec{The Chern-Simons Term in 2 + 1 Dimensions}

Before moving on, let us take a closer look at the proposed $2+1$ dimensional 
Chern-Simons term in the case corresponding to matrix theory on 
$\S^1/\Z_2\times\S^1$.  Chern-Simons terms in $2+1$ dimensions change their 
sign under parity, and it is somewhat unusual to suggest their presence in 
a theory which is projected to the parity-invariant sector.  In the case at 
hand, this question is resolved as follows:  the Chern-Simons term manages 
to be parity-even, because its ``coupling constant'' is itself odd under 
parity!  Indeed, the Chern-Simons coupling is given by the RR field strength, 
which in $2+1$ dimensions is a zero-form.  It is constant everywhere outside 
the singular locus $V_\alpha$ (and away from the D8-branes at 
$S_{\tilde\alpha}$).  However, at $V_\alpha$ (as well as $S_{\tilde\alpha}$) 
it develops a step-function singularity, because $V_\alpha$ (and 
$S_{\tilde\alpha}$) serve as sources in its Bianchi identity.  Ignoring the 
D8-branes for the moment, \eecstone\ in $2+1$ dimensions close to the boundary 
can be written as 
\eqn\eecsthree{\int G\,\tr\,\omega_3=k\int\varepsilon(\sigma_2)\,\tr\,
\omega_3,}
where $G\equiv k\varepsilon(\sigma_2)$; this indeed makes the Chern-Simons 
term even under parity.  

Notice that this is somewhat reminiscent of certain Chern-Simons terms in 
$4+1$ dimensions, encountered in a closely related context on the 
world-volume of five-brane probes in Type IA vacua \fivecs ; it also exhibits 
some similarities with the mechanism that ensures parity invariance in 
effective M-theory in eleven dimensions despite the presence of its 
Chern-Simons self-interaction term \ewflux .  

\newsec{D8-Branes from Matrices}

Although the Chern-Simons couplings to a non-trivial RR background 
solve the anomaly cancellation problem in the matrix theory description of 
the potentially anomalous configurations, the explicit presence of a 
closed-string background is not entirely satisfactory.  One of the 
lessons we have learned in D-brane physics \refs{\dkps,\mdrev} is that the 
long-distance effects of gravity -- and more generally, effects connected 
with a closed-string exchange -- are reinterpreted at substringy scales as 
open-string effects.  While it is conceivable that the matrix theory 
description of certain vacua may require explicit couplings to closed 
string backgrounds of the type we encountered in section~3, one might be 
tempted to interpret such couplings as part of an effective description of a 
more microscopic mechanism, with the supergravity background being generated 
when certain heavy matrix modes are integrated out.  In this section we will 
suggest that these extra degrees of freedom are related to the microscopic 
description of D8-branes in matrix theory.  

So far in this paper, the extra fermions $\chi$ have been the only indication 
that the $\S^1/\Z_2\times\T^d$ vacuum of matrix theory contains D8-branes.%
\foot{As in the rest of the paper, we will refer to the longitudinally wrapped 
9-branes of M-theory as ``D8-branes'' for short.}
We have seen that the fermions $\chi$ are not needed to cancel local gauge 
anomalies at the fixed-point planes $V_\alpha$; generically, they are not 
even located at $V_\alpha$, and should presumably be thought of as localized 
bulk degrees of freedom.  These observations suggest that the corresponding 
D8-branes could have an intrinsic description in matrix theory, as localized 
bulk configurations in the orbifold Yang-Mills theory.  In this section we 
present evidence that this is indeed the case, and demonstrate that D8-branes 
appear as certain topologically non-trivial configurations in the Yang-Mills 
theory.  

\subsec{D8-Branes in Matrix Theory}

In the previous section, we have seen that in order to ensure local 
cancellation of anomalies, the surfaces $S_{\tilde\alpha}$ that support the 
chiral fermions $\chi$ had to be postulated to carry one unit of the D8-brane 
charge.  We want to replace this argument by a more microscopic picture, and 
interpret the fermion as a zero mode in a Yang-Mills background which itself 
carries the corresponding charge.  Consider Yang-Mills configurations with 
non-zero fourth Chern character, 
\eqn\eeinsta{{\it ch}_4(F)=\frac{1}{4!(2\pi)^4}\int\tr\left(F\wedge F\wedge F
\wedge F\right).}
Such configurations couple naturally to the corresponding RR field $C'$, and 
carry a non-zero D8-brane charge given by the value of ${\it ch}_4(F)$.  The 
configuration with the minimum unit of this charge
\eqn\eecnser{\frac{1}{4!(2\pi)^4}\int\tr\left(F\wedge F\wedge F\wedge F\right)
=1,}
 is a natural candidate for the matrix theory description of one D8-brane.%
\foot{This D8-brane will in general carry stacks of lower-dimensional branes 
in its world-volume.  In order to have a configuration with only the D8-brane 
charge, we would have to make sure that the corresponding lower-dimensional 
cohomology classes vanish.}
A virtually identical conjecture has appeared in a broader context in 
\brafrom ; our analysis, based on arguments motivated by the anomaly 
cancellation mechanism of section~3, provides further support for this 
conjecture.  

In the decompactified case, i.e.\ in matrix theory on $\S^1/\Z_2$ described 
by $1+1$ dimensional Yang-Mills theory, the configurations of matrices that 
carry non-zero ${\it ch}_4(F)$ are translated with the use of \eexder\ to 
configurations of matrices with the corresponding non-zero value of 
\eqn\eeightabs{\tr\;\left([X,X]\wedge [X,X]\wedge [X,X]\wedge [X,X]\right).}
(Here the wedge products indicate that the space-time indices are contracted 
with the completely antisymmetric tensor in the eight transverse dimensions.) 
This configuration carries infinite energy, as expected, since it describes a 
configuration of branes with infinite volume.  

How do we construct explicit solutions of the Yang-Mills equations of motion 
that carry non-zero D8-brane charge \eeinsta ?  One might be tempted to 
consider some kind of self-duality conditions in eight dimensions, of the 
type
\eqn\eesdeight{F_{\mu\nu}=T_{\mu\nu\sigma\rho}F^{\sigma\rho},}
with $T^{\mu\nu\sigma\rho}$ a certain fixed four-form in eight dimensions.  
A large literature is devoted to this subject (see e.g.\ \eightinst ).  
Self-duality in eight dimensions indeed exhibits some remarkable properties: 
group-theoretical classifications of possible four-forms $T^{\mu\nu\sigma
\rho}$ have been presented, and an analogy of the ADHM construction has been 
shown to exist in some cases; the algebra of octonions plays a prominent role 
in some of these constructions.  In general, however, 
the corresponding self-dual Yang-Mills configurations break more supersymmetry 
than we want for the present purpose.  While self-duality in eight dimensions 
as discussed in \eightinst\ might still correspond to interesting brane 
configurations in matrix theory, we will resort to other means that will allow 
us to construct a D8-brane configuration in matrix theory. 

To construct a Yang-Mills configuration with non-zero D8-brane charge 
\eeinsta , we will use lower-dimensional brane configurations.  Decompose 
the eight-torus as the product of two four-tori, $\T_1\times\T_2$, and 
consider two four-dimensional $SU(2)$ Yang-Mills instantons $\CA^{(1)}$ and 
$\CA^{(2)}$ with instanton number one on $\T_1$ and $\T_2$ respectively.  Set 
\eqn\eecombinst{\tilde A=\CA^{(1)}+\CA^{(2)};}
this defines an $SU(2)$ gauge potential on $\T_1\times\T_2$ that carries one 
unit of the D8-brane charge \eeinsta .  As in \brafrom , this configuration 
on $\T_1\times\T_2$ has no gauge invariant moduli; when translational 
invariance is broken, for example by the orbifold $\Z_2$ action on the torus, 
the configuration will have a gauge invariant modulus, corresponding to the 
relative location of the instanton with respect to the orbifold 
singularities.  

In addition to the D8-brane charge, the configuration \eecombinst\ will carry 
D4-brane charges, due to the presence of D4-branes represented by the 
individual instantons $\CA^{(1)}$ and $\CA^{(2)}$ in the world-volume of 
the D8-brane.  We have been unable to find a configuration which would only 
carry the D8-brane charge while preserving the amount of supersymmetry 
required by the longitudinally wrapped 9-brane of M-theory on 
$\S^1/\Z_2\times\T^d$. 

Modulo this fact, we are now ready to describe a system of D0-branes in the 
presence of the D8-brane.  Assume that the Yang-Mills configuration carrying 
one unit of the 
D8-brane charge \eeinsta\ is given by a gauge potential $\tilde A$ in an 
$SU(N_0)$ subgroup of the full gauge group; for simplicity, we will refer to 
$\tilde A$ as the ``instanton.''  (As an example, one can consider 
\eecombinst , in which case $N_0=2$ and the ``instanton'' is constructed from 
the two four-dimensional $SU(2)$ instantons localized in dimensions 
$2\ldots 5$ and $6\ldots 9$ respectively.)  To describe the background with 
$k$ D8-branes, we select $k$ mutually commuting $U(N_0)$ subgroups in the 
Yang-Mills gauge group, and consider the configurations with one unit of 
D8-brane charge in each of the $U(N_0)$ sectors.  The choice of the $k$ 
subgroups breaks the gauge symmetry of matrix theory to 
\eqn\eebrgr{U(N)\times\underbrace{U(N_0)\times\ldots\times U(N_0)}_k,}
which is further broken in the $U(N_0)$ sectors by the $SU(N_0)$ gauge 
backgrounds $\tilde A$.  In the background of $k$ instantons, we will 
decompose the Yang-Mills gauge field as 
\eqn\eematinst{\pmatrix{A&B^1&\ldots&B^k\cr
B^{\dagger 1}&\tilde A^{(1)}+C^{11}&\ldots&C^{1k}\cr
\vdots&\vdots&\ddots&\vdots\cr
B^{\dagger k}&C^{k1}&\ldots&A^{(k)}+C^{kk}\cr}}
(and similarly for the fermions).  In \eematinst , $\tilde A^{(p)}$ is 
the instanton gauge field bacground in the $p$-th copy of $U(N_0)$.  

According to our conjecture, this configuration describes $N$ D0-branes 
in the presence of $k$ D8-branes (possibly with lower-dimensional branes 
in their world-volumes if its lower-dimensional Chern numbers are non-zero 
\brafrom ).  In this scenario, D8-branes appear as composites of ``partonic 
D0-branes,'' in accord with the original philosophy of the BFSS proposal 
\bfss .  

\subsec{0-8 Strings and 8-8 Strings from Collective Matrix Coordinates}

In the instanton background \eematinst , matrix theory contains the adjoint 
$U(N)$ degrees of freedom (in the block denoted by $A$ in \eematinst ), plus 
other degrees of freedom due to the presence of the instanton background.  
We stress that all these degrees of freedom are a part of the original 
adjoint large-$N$ matrices; the choice of the subgroup that contains the 
instanton background breaks the original large-$N$ gauge group to \eebrgr .  
It is natural to interpret the $U(N)$ group of \eebrgr\ in the large-$N$ limit 
as the matrix theory gauge group, and derive an effective matrix theory in 
the presence of the D8-brane degrees of freedom.  In this effective theory, 
in addition to the adjoint $U(N)$ fields, one would keep only the zero modes 
and integrate out the non-zero modes of the matrix elements that are not in 
the adjoint of $U(N)$, thus generating an effective description of the theory 
in terms of $N\rightarrow\infty$ D0-branes that are not ``partonic 
components'' of the D8-branes, plus extra degrees of freedom from the zero 
modes.  

With this picture in mind, consider the off-diagonal blocks in \eematinst\ 
(and their superpartners).  

\item{(1)}  
The off-diagonal matrices $B$ carry the fundamental representation of the 
$U(N)$ gauge group of the $N$ D0-branes that do not appear as partons in the 
D8-branes.  Microscopically, the matrix elements in $B$ (and their 
superpartners) represent 0-0 strings stretching between a given D0-brane and 
a ``parton'' within the D8-brane; effectively, they should give rise to the 
0-8 strings.  

\item{(2)}  The $C^{\ell m}$ matrices in \eematinst\ are $U(N)$ singlets, and 
for $\ell\neq m$ carry the fundamental of the $\ell$-th and $m$-th copy of 
$U(N_0)$.  Microscopically, they represent open strings stretching between two 
``partons'' in the corresponding two D8-branes, and should therefore be 
interpreted as the matrix-theory description of the 8-8 strings.  

We expect the fermions $\chi$ to appear as the low-energy collective 
excitations in the $B$ sector of \eematinst .  To check this, we will study 
the effective $U(N)$ dynamics in the presence of a non-trivial instanton gauge 
field $\tilde A$.  To be specific, we will again take $\tilde A$ to be the 
instanton of \eecombinst .  Each four-dimensional instanton $\CA$ in 
\eecombinst\ has a chiral fermion zero-mode in the fundamental representation 
of $U(N)$; hence, the eight-dimensional instanton \eecombinst\ will have a 
chiral fermion zero mode $\chi_0$, satisfying 
\eqn\eeeizerom{\Gamma^\mu D_\mu (\tilde A)\chi_0\equiv D(\CA^{(1)})\chi_0+
D(\CA^{(2)})\chi_0=0.}
(Here $D(\CA)$ is the four-dimensional Dirac operator in the instanton 
background $\CA$.)  
This zero mode is chiral in eight dimensions; in the two dimensions 
$(\sigma^0,\sigma^1)$ transverse to the instanton, it gives rise to a chiral 
$1+1$ dimensional field $\chi_0$ in the fundamental representation of $U(N)$.  
In the presence of $k$ D8-branes, one gets $k$ chiral $1+1$ dimensional 
fermions in the fundamental of $U(N)$, and described by the effective action 
\eqn\eefefefe{\int_{S_{\tilde\alpha}}d^2\sigma\;\tr\;\bar\chi_0(\p_++A_+)
\chi_0,}
precisely as expected on the basis of the suggested relation to the 0-8 
strings.  The $1+1$ dimensional surfaces $S_{\tilde\alpha}$ in \eefefefe\ are 
at the core of the corresponding instantons.  

Similarly, consider the $C$ sector.  According to our conjecture, the zero 
modes should correspond to the massless modes of the 8-8 strings.  
Masses of the lowest 8-8 string modes are proportional to the distance 
between the corresponding D8-branes (and depend appropriately on the Wilson 
lines, should those be present in the vacuum).  In particular, when two or 
more branes coincide, the world-volume gauge symmetry is enhanced.  Now we 
will show that the instanton background \eematinst\ contains zero modes that 
exhibit this enhanced symmetry behavior.  
In the general instanton background, the off-diagonal Yang-Mills 
gauge fields gain masses, and the gauge symmetry \eebrgr\ is broken to 
\eqn\eefubr{U(N)\times\underbrace{U(1)\times\ldots\times U(1)}_k.}
The $k$ copies of $U(1)$ are unbroken as the instantons take values in 
$SU(N_0)$.  However, if the locations of two instantons coincide in the 
$\sigma$ space, a new massless mode appears, as we can now rotate the two 
instantons as rigid objects in $U(2N_0)$.  Together with the unbroken 
$U(1)$'s, this generates the expected $U(2)$.  If $k$ instantons coincide, 
this gives rise to bosonic zero modes 
\eqn\eefuzm{\pmatrix{0&0&\ldots&0\cr
0&a_{11}I_{N_0}&\ldots&a_{1k}I_{N_0}\cr
\vdots&\vdots&\ddots&\vdots\cr
0&a_{k1}I_{N_0}&\ldots&a_{kk}I_{N_0}\cr},}
where $I_{N_0}$ is the $N_0\times N_0$ unit matrix, and 
\eqn\eeumatr{\pmatrix{a_{11}&\ldots&a_{1k}\cr
\vdots&\ddots&\vdots\cr
a_{k1}&\ldots&a_{kk}\cr}}
is in the adjoint representation of $U(k)$.  The $U(k)$ transformations can 
depend on the dimensions $\sigma^\mu$ ($\mu=0,1$) transverse to the instanton, 
and we recover the bosonic sector of the gauge theory on the world-volume of 
the D8-brane.%
\foot{More precisely, on the two-dimensional locus $S_{\tilde\alpha}$ which 
corresponds to the D8-brane location in the Yang-Mills parameter space.}
This is in accord with the suggested interpretation of these zero modes as the 
massless 8-8 string modes.  

In this section, we have suggested that D8-branes in matrix theory are 
described by non-trivial matrix configurations.  In this scenario, 
the massless modes of 0-8 strings and 8-8 strings emerge as the zero modes in 
the corresponding background, and do not have to be artificially added into 
the theory.  We can integrate out the off-digonal non-zero modes, thus 
obtaining an effective description in terms of the D0-branes and the 
extra degrees of freedom from the zero modes.  We expect that this should 
reproduce the Chern-Simons couplings with effective supergravity backgrounds, 
in the spirit of \dkps .  This is indeed plausible, since the non-zero modes 
include degrees of freedom that can be interpreted as massive 0-8 (and 8-8 
strings), which are known to be responsible for the effective supergravity 
background at large distances.  

Given that the D8-branes may have an intrinsic description in matrix theory, 
one wonders whether the orientifold planes can themselves be described in an 
intrinsic way.  They serve as sources of the RR background, which is again 
indicative of an effective theory.  We have nothing to say about this in this 
paper, except for noting that such a description may not be completely 
unexpected.  Indeed, Sen has shown \fsen\ that in certain orientifold 
compactifications of Type IIB string theory, the orientifold planes split 
non-perturbatively in the string coupling constant into a pair of D-branes.  
In the matrix theory context, such phenomena would indicate that the theory 
may not respect the orbifold procedure assumed as a part of the kinematics 
in our construction; the full Yang-Mills theory may no longer be an 
orbifold theory non-perturbatively.  

\newsec{Orbifold Yang-Mills Dynamics from Matrix Theory}

Having seen in sections~2 and~3 how the orbifold Yang-Mills gauge theories 
with eight supercharges appear in matrix theory on $\S^1/\Z_2\times\T^d$, 
we will now leave the semi-classical regime, and 
consider the full dynamics of the orbifold theories.  In this section, 
following the strategy of \fhrs , we will briefly discuss how the assumed 
correspondence with the heterotic string theory on tori leads to predictions 
about the non-perturbative dynamics of the corresponding orbifold Yang-Mills 
theories.  As in \fhrs , the expected properties of string theory suggest the 
existence of a multitude of fixed points in various dimensions.  

To keep our discussion simple, we will avoid non-trivial supergravity 
backgrounds by assuming the symmetric distribution of D8-branes among the 
components of the singular locus.  

\subsec{Yang-Mills Dynamics on $\S^1\times\S^1/\Z_2$}

When the size of the space-time cylinder shrinks to zero volume, matrix theory 
on $\S^1/\Z_2\times\S^1$ is supposed to reproduce the Type I and heterotic 
$SO(32)$ string theory \hw\ in the light cone gauge.  In particular, since 
these string theories exhibit $\CN=1$ super Poincar\'e invariance in ten 
dimensions, our matrix theory description should recover the full $SO(8)$ 
Lorentz invariance in the transverse space-time dimensions.  In the orbifold 
Yang-Mills theory, only an $SO(7)$ subgroup is manifest.  This situation 
has been analyzed in the related case with maximum supersymmetry 
\refs{\sethis,\strfrom}, and it has been argued that in the appropriate 
limit, the theory grows another macroscopic space-time dimension.  The 
space-like torus of the $2+1$ dimensional Yang-Mills theory can carry a 
non-zero magnetic flux, and its quanta behave as KK modes along an extra 
space-time dimension of radius
\eqn\eeextrastrad{R'=(2\pi)^3\frac{\ell_{11}^3}{R_1R_2}.}

In the orbifold case, the torus becomes a cylinder.  The first interesting 
question to be asked is what is the behavior of the dynamically generated 
space-time dimension under the $\Z_2$ orbifold symmetry.  Since the KK modes 
along the extra dimension correspond to the quanta of magnetic flux through 
that cylinder, they are even under the $\Z_2$ symmetry.  Hence, the extra 
dimension is also even under the orbifold group, and the arguments 
\refs{\sethis,\strfrom} for the enhanced $SO(8)$ invariance in the IR carry 
over to the orbifold theory.  

The correspondence with the heterotic and Type I $SO(32)$ string theory 
thus predicts the existence of a non-trivial infrared fixed point with 
eight supercharges in the Yang-Mills gauge theory on the $2+1$-dimensional 
orbifold.  The supercharges are in the ${\bf 8}_s$ of the global $SO(8)$ of R 
symmetries.  The fixed point is associated with the superconformal group of 
the half-infinite $2+1$ manifold $\R^2\times\R/\Z_2$, with supercharges that 
respect chiral boundary conditions.  

\subsec{Yang-Mills Dynamics on $\S^1\times\T^2/\Z_2$}

In $3+1$ dimensions, the dimensionless Yang-Mills coupling is given in terms 
of the space-time compactification parameters by
\eqn\eecoupfodi{g^2=(2\pi)^4\frac{\ell_{11}^3}{R_1R_2R_3}.}
The theory contains states with non-zero magnetic flux through the three 
two-dimensional cycles of the orbifold.  As the volume of the space-time 
$\S^1/\Z_2\times\T^2$ shrinks to zero, these modes become light, and can be 
naturally interpreted as KK modes along three macroscopic dimensions in 
space-time \refs{\lsdual,\grtdual,\fhrs}.  The magnetic fluxes through the 
$\S^1/\Z_2\times\S^1$ two-cycles are even under the $\Z_2$ orbifold group, 
and therefore correspond to space-time dimensions that are also even.  On 
the other hand, the flux through the $\T^2/\Z_2$ cycle changes its sign under 
the $\Z_2$ action, and represents a KK mode along a space-time dimension which 
changes its orientation under the $\Z_2$.  As the volume of the original 
$\S^1/\Z_2\times\T^2$ shrinks to zero, the macroscopic space-time again 
corresponds to M-theory compactified on $\S^1/\Z_2\times\T^2$.  As in the 
case with maximum supersymmetry \fhrs , these two compactifications are 
related in the Yang-Mills theory by
\eqn\eeemd{g^2\rightarrow\frac{(2\pi)^2}{g^2}.}
The space-time T-duality between the two heterotic string compactifications 
predicts the existence of an electric-magnetic duality between the 
corresponding orbifold Yang-Mills theories with eight supercharges.  

In principle, one would want to reconstruct in matrix theory the full 
U-duality group of the heterotic string vacua on $\T^2$.  A full analysis 
would require a more detailed understanding of orbifold Yang-Mills dynamics 
in the presence of generic space-time Wilson lines, and is clearly beyond the 
scope of this paper.  

\subsec{Yang-Mills Dynamics on $\S^1\times\T^3/\Z_2$}

Before orbifolding by $\Z_2$, this case corresponds to matrix theory on a 
four-torus.  The gauge coupling is 
\eqn\eecoupfidi{g^2=(2\pi)^6\frac{\ell_{11}^6}{\tilde RR_1R_2R_3R_4};}
the effective dimensionless coupling equals 
\eqn\eecoupfill{\tilde g^2\equiv\frac{g^2}{(\rho_1\rho_2\rho_3\rho_4)^{1/4}}
=(2\pi)^4\frac{\ell_{11}^3}{(R_1R_2R_3R_4)^{3/4}}.}
As the space-time four-torus shrinks uniformly to zero volume, the gauge 
theory is accordingly expected to flow to a UV fixed point \fhrs .  It was 
pointed out, however, that there are no known fixed points with the expected 
properties in $4+1$ dimensions.  Instead, the proper description of the UV 
fixed point seems to be in terms of a $(0,2)$ supersymmetric non-trivial 
fixed point in $5+1$ dimensions \refs{\rozali,\fhrs,\berro}.  

In terms of the $4+1$ dimensional theory, the extra dimension appears because 
the theory contains states that become light in the appropriate limit 
\rozali .  These states are given by marginal bound states of 4-dimensional 
Yang-Mills instantons, which represent solitonic particles in the $4+1$ 
dimensional theory.  The state with instanton number $Q$ carries energy
\eqn\eeoneinsen{E=\frac{(2\pi)^2Q}{g^2},}
which suggests that $Q$ can be interpreted as the wave number in an extra 
dimension of radius
\eqn\eewsdimrad{\rho'=(2\pi)^5\frac{\ell_{11}^6}{\tilde RR_1R_2R_3R_4}.}

Upon orbifolding one space-time dimension, we obtain a Yang-Mills theory on 
$\S^1\times\T^3/\Z_2$.  What is the behavior of the extra dimension under the 
orbifold group $\Z_2$?  Since the orbifold group reflects three out of four 
dimensions of the torus, the instanton number is odd, therefore the extra 
dimension is odd under the $\Z_2$ orbifold group. The total $\Z_2$ acts on 
the product of the $\T^3$ and the extra dimension, and the theory becomes a 
$5+1$ dimensional field theory on the corresponding orbifold.  String 
dynamics again suggests the existence of a non-trivial UV fixed point in 
this six-dimensional orbifold field theory with eight supercharges.  

\bigskip\bigskip
\centerline{\bf Acknowledgement}
\medskip
These results were presented at the Second Trieste Conference on Duality 
Symmetries in String Theory at the ICTP, Trieste, April 1-4, 1997.  I would 
like to thank the organizers for their hospitality and for creating a 
stimulating atmosphere during the meeting.  

\listrefs
\end